\PassOptionsToPackage{unicode}{hyperref}
\PassOptionsToPackage{hyphens}{url}
\PassOptionsToPackage{dvipsnames,svgnames,x11names}{xcolor}
\documentclass[
]{article}
\usepackage{amsmath,amssymb}
\usepackage{lmodern}
\usepackage{iftex}
\usepackage[T1]{fontenc}
\usepackage[utf8]{inputenc}
\usepackage{textcomp} 
\IfFileExists{upquote.sty}{\usepackage{upquote}}{}
\IfFileExists{microtype.sty}{
  \usepackage[]{microtype}
  \UseMicrotypeSet[protrusion]{basicmath} 
}{}
\usepackage[margin=1in]{geometry}{}
\makeatletter
\@ifundefined{KOMAClassName}{
  \IfFileExists{parskip.sty}{%
    \usepackage{parskip}
  }{
    \setlength{\parindent}{0pt}
    \setlength{\parskip}{6pt plus 2pt minus 1pt}}
}{
  \KOMAoptions{parskip=half}}
\makeatother
\usepackage{xcolor}
\NewDocumentCommand\citeproctext{}{}
\NewDocumentCommand\citeproc{mm}{%
  \begingroup\def\citeproctext{#2}\cite{#1}\endgroup}
\makeatletter
 \let\@cite@ofmt\@firstofone
 \def\@biblabel#1{}
 \def\@cite#1#2{{#1\if@tempswa , #2\fi}}
\makeatother
\newlength{\cslhangindent}
\newlength{\csllabelwidth}
\newenvironment{CSLReferences}[2] 
 {\begin{list}{}{%
  \setlength{\itemindent}{0pt}
  \setlength{\leftmargin}{0pt}
  \setlength{\parsep}{0pt}
  \ifodd #1
   \setlength{\leftmargin}{\cslhangindent}
   \setlength{\itemindent}{-1\cslhangindent}
  \fi
  \setlength{\itemsep}{#2\baselineskip}}}
 {\end{list}}
\usepackage{calc}

\ifLuaTeX
\usepackage[bidi=basic]{babel}
\else
\usepackage[bidi=default]{babel}
\fi
\babelprovide[main,import]{american}

\def\languageshorthands#1{}
\ifLuaTeX
  \usepackage{selnolig}  
\fi
\IfFileExists{bookmark.sty}{\usepackage{bookmark}}{\usepackage{hyperref}}
\IfFileExists{xurl.sty}{\usepackage{xurl}}{} 
\hypersetup{
  pdftitle={MacroEnergy.jl: A large-scale multi-sector energy system
framework},
  pdfauthor={Ruaridh Macdonald, Filippo Pecci, Luca Bonaldo, Jun Wen
Law, Yu Weng, Dharik Mallapragada, Jesse Jenkins},
  pdflang={en-US},
  colorlinks=true,
  linkcolor={Maroon},
  filecolor={Maroon},
  citecolor={Blue},
  urlcolor={Blue},
  pdfcreator={LaTeX via pandoc}}

\title{MacroEnergy.jl: A large-scale multi-sector energy system
framework}

\definecolor{c53baa1}{RGB}{83,186,161}
\definecolor{c202826}{RGB}{32,40,38}

\usepackage[affil-it]{authblk}
\usepackage{orcidlink}
\author[1%
  \ensuremath\mathparagraph]{Ruaridh Macdonald%
    \,\orcidlink{0000-0001-9034-6635}\,%
    }
\author[2%
  ]{Filippo Pecci%
    \,\orcidlink{0000-0003-3200-0892}\,%
    }
\author[3%
  ]{Luca Bonaldo%
    \,\orcidlink{0009-0000-0650-0266}\,%
    }
\author[1%
  ]{Jun Wen Law%
    \,\orcidlink{0009-0001-8766-3100}\,%
    }
\author[1%
  ]{Yu Weng%
    \,\orcidlink{0000-0003-3958-1546}\,%
    }
\author[4%
  ]{Dharik Mallapragada%
    \,\orcidlink{0000-0002-0330-0063}\,%
    }
\author[3%
  ]{Jesse Jenkins%
    \,\orcidlink{0000-0002-9670-7793}\,%
    }

\affil[1]{Massachusetts Institute of Technology, USA%
  }
\affil[2]{RFF-CMCC European Institute on Economics and the Environment,
Italy%
  }
\affil[3]{Princeton University, USA%
  }
\affil[4]{New York University, USA%
  }
\affil[$\mathparagraph$]{Corresponding author: Ruaridh Macdonald%
}
\date{24 October 2025}

\begin{document}
\maketitle

\section{Summary}\label{summary}

MacroEnergy.jl (aka Macro) is an open-source framework for multi-sector
capacity expansion modeling and analysis of macro-energy
systems(\citeproc{ref-levi2019macro}{Levi et al., 2019}). It is written
in Julia (\citeproc{ref-bezanson2017julia}{Bezanson et al., 2017}) and
uses the JuMP (\citeproc{ref-dunning2017jump}{Dunning et al., 2017})
package to interface with a wide range of mathematical solvers. It
enables researchers and practitioners to design and analyze energy and
industrial systems that span electricity, fuels, bioenergy, steel,
chemicals, and other sectors. The framework is organized around a small
set of sector-agnostic components that can be combined into flexible
graph structures, making it straightforward to extend to new
technologies, policies, and commodities. Its companion packages support
decomposition methods and other advanced techniques, allowing users to
scale models across fine temporal and spatial resolutions.
MacroEnergy.jl provides a versatile platform for studying energy
transitions at the detail and scale demanded by modern research and
policy.

\section{Statement of Need}\label{statement-of-need}

The increasing complexity of energy systems necessitates advanced
modeling tools to support decision-making in infrastructure planning,
R\&D decisions and policy design. This complexity comes from the
challenge of ensuring the reliability of grids with large amounts of
renewable generation and storage, increased coupling and electrification
of energy-intensive sectors, greater diversity in the technologies and
policies being deployed, and many other factors.

Capacity expansion modelling frameworks have improved substantially in
recent years. A wider range of problems can now be solved thanks to
improvements in the underlying formulations and solvers while access to
richer data sources has enabled more realistic representations of
resources, weather and demand. Looking ahead, further improvements are
on the horizon, including non-linear technology formulations that
capture richer trade-offs (\citeproc{ref-falth2023trade}{Fälth et al.,
2023}; \citeproc{ref-heo2024effects}{Heo \& Macdonald, 2024};
\citeproc{ref-levin2023energy}{Levin et al., 2023}), tighter integration
with integrated assessment models and other tools
(\citeproc{ref-gong2023bidirectional}{Gong et al., 2023};
\citeproc{ref-gotske2025first}{Gøtske et al., 2025};
\citeproc{ref-odenweller2025remind}{Odenweller et al., 2025}), and novel
approaches to scaling up problem size
(\citeproc{ref-liu2024generalized}{Liu et al., 2024};
\citeproc{ref-parolin2025sectoral}{Parolin et al., 2025};
\citeproc{ref-pecci2025regularized}{Pecci \& Jenkins, 2025}).

There has also been some convergence in the design and capabilities of
modelling frameworks as the field comes to understand what is required
to produce robust, policy-relevant results. Recent studies suggest that
capacity expansion models must consider decades of operational data
(\citeproc{ref-ruggles2024planning}{Ruggles et al., 2024};
\citeproc{ref-ruhnau2022storage}{Ruhnau \& Qvist, 2022}), may require
temporal resolution as fine as five minutes
(\citeproc{ref-levin2024high}{Levin et al., 2024};
\citeproc{ref-mallapragada2018impact}{Mallapragada et al., 2018}), and
should capture spatial heterogeneity at the county level
(\citeproc{ref-frysztacki2023inverse}{Frysztacki et al., 2023};
\citeproc{ref-krishnan2016evaluating}{Krishnan \& Cole, 2016};
\citeproc{ref-qiu2024decarbonized}{Qiu et al., 2024};
\citeproc{ref-serpe2025importance}{Serpe et al., 2025}). In addition,
they must be able to represent a wide variety of coupled sectors as the
majority of emission reductions will come from outside the electricity
sector. Electricity-centric frameworks; such as PyPSA
(\citeproc{ref-brown2017pypsa}{T. Brown et al., 2017}), GenX
(\citeproc{ref-jenkins2017enhanced}{Jenkins \& Sepulveda, 2017}),
Calliope (\citeproc{ref-pfenninger2018calliope}{Pfenninger \& Pickering,
2018}), and others (\citeproc{ref-blair2014system}{Blair et al., 2014};
\citeproc{ref-Brown_Regional_Energy_Deployment}{P. Brown et al., n.d.};
\citeproc{ref-he2024dolphyn}{He et al., 2024};
\citeproc{ref-howells2011osemosys}{Howells et al., 2011}); developed the
computational capabilities needed to optimize grids over long time
series of hourly or sub-hourly data in order to properly incorporate
variable renewable energy generation and storage. In recent years,
several have begun to extending their frameworks to include other
sectors, such as hydrogen, fuels, and industrial processes. On the other
hand, economy-wide models; such as TIMES
(\citeproc{ref-loulou2005documentation}{Loulou et al., 2005}), TEMOA
(\citeproc{ref-hunter2013modeling}{Hunter et al., 2013}) and others;
have long been able to represent multiple sectors though the use of
flexible graph-based structures. However, they do not have the
computational performance required to include long, high-resolution time
series.

Extending existing models to new sectors or to dramatically improve
performance often requires rewriting core routines or layering new
modules on top. This complicates validation, obscures interactions
across the system, and leaves the codebase hard to maintain. In the
authors' experience from previous development, the frameworks remain
architectured around their original sectors, making it problematic to
exclude those sectors and quickly increasing the difficulty and time
required to add new features.

MacroEnergy.jl was designed to overcome these limitations. Its
architecture is based on a small set of sector-agnostic components that
can be combined into graphs to represent networks, technologies, and
policies in any sector. Features are largely independent of one another,
allowing users to focus on how best to represent their technology or
policy of interest instead of working around the existing code.

MacroEnergy.jl is also designed from the ground-up to scale to large,
multi-sector problems. Modeling across coupled sectors greatly increases
runtimes, often making problems intractable
(\citeproc{ref-parolin2025sectoral}{Parolin et al., 2025}). Techniques
such as model compression and the use of representative periods can ease
the computational burden, but eventually large-scale models reach the
limits of what can be solved on a single computing node. To scale
further, methods which allow models to be solved across computing
clusters are essential. MacroEnergy.jl was designed with these
challenges in mind. Its data structures and graph-based representation
of energy systems enable sectoral, temporal and spatial decompositions
by default. It also includes a suite of companion packages, which
provide advanced decomposition algorithms
(\citeproc{ref-pecci2025MacroEnergySolvers}{Pecci et al., 2025}),
automatic model scaling
(\citeproc{ref-macdonald2024MacroEnergyScaling}{Macdonald, 2024}), and
example systems
(\citeproc{ref-macdonald2025MacroEnergyExamples}{Macdonald et al.,
2025}). Other companion packages are under development. These will
provide representative period selection and other tools to enhance
MacroEnergy.jl. MacroEnergy.jl and its companion packages are registered
Julia packages and are freely available on GitHub or through the Julia
package manager.

\section{Use Cases}\label{use-cases}

MacroEnergy.jl can be used to optimize the design and operation of
energy and industrial systems, investigate the value of new technologies
or polices, optimize investments in an energy system over multiple
years, and many other tasks. It is being used for several ongoing
investigations of regional energy systems, including as part of the
Net-Zero X Global Initiative - a research consortium involving top
research institutions around the world developing shared modeling
methods and completing detailed, actionable country-specific studies
supporting net-zero transitions.

The framework was designed with three user profiles in mind. Where
possible, we have passed modelling complexity upstream to developers, so
that most users can build and run models faster and with less coding
knowledge.

\begin{itemize}
\item
  Users: Want to create and optimize a real-world system using
  MacroEnergy.jl. They should be able to do this with little or no
  coding, and without knowledge of MacroEnergy.jl's components or
  internal structure.
\item
  Modelers: Want to add new assets, sectors, or public policies to
  MacroEnergy.jl. They will need to be able to code in Julia and
  understand some of MacroEnergy.jl's components, but they do not
  require knowledge of its internal structure or underlying packages.
\item
  Developers: Want to change or add new features, model formulations or
  constraints to MacroEnergy.jl. They will require detailed knowledge of
  MacroEnergy.jl's components, internal structure, and underlying
  packages.
\end{itemize}

\section{Structure}\label{structure}

MacroEnergy.jl models are made up of four core components which are used
to describe the production, transport, storage and consumption of
various commodities. The components can be connected into multi-sectoral
networks of commodities. They are commodity-agnostic so can be used for
any flow of a good, energy, etc. While we believe MacroEnergy.jl will
most often be used to study energy systems, commodities can also be
data, money, or more abstract flows.

The four core components are:

\begin{enumerate}
\def\labelenumi{\arabic{enumi}.}
\item
  Edges: describe and constrain the flow of a commodity
\item
  Nodes: balance flows of one commodity and allow for exogenous flows
  into and out of a model. These can be used to represent exogenous
  demand or supply of a commodity.
\item
  Storage: allow for a commodity to be stored over time.
\item
  Transformations: allow for the conversion of one commodity into
  another by balancing flows of one or more commodities.
\end{enumerate}

These four core components can be used directly to build models but most
users will find it easier to combine them into Assets and Locations.
Assets are collections of components that represent real-world
infrastructure such as power plants, industrial facilities, transmission
lines, etc. For example, a water electrolyzer asset would include edges
for electricity and water inputs and hydrogen output, and a
transformation to conver between them. Locations are collections of
Nodes which represent physical places where assets are situated and
commodities can be transported between. While Edges can only connect to
Nodes of the same Commodity, Locations are an abstraction that
simplifies the user-input required to connect different commodities
across physical places. Together, Assets and Locations allow for models
to be truer to life and easier to analyze.

Assets and Locations in turn form Systems which represent an energy
and/or industrial system. Most often, each System will be optimized
separately given a user-defined operating period. Several Systems can be
combined into a Case. Cases can be used for multi-stage capacity
expansion models, rolling-horizon optimization, sensitivity studies, and
other work requiring multiple snapshots or versions of an energy system.
MacroEnergy.jl can automatically manage the running of these different
Cases for users, either directly or in combination with
MacroEnergySolver.jl package.

\section{Acknowledgements}\label{acknowledgements}

The development of MacroEnergy.jl was funded by the Schmidt Sciences
Foundation. This publication was based (fully or partially) upon work
supported by the U.S. Department of Energy's Office of Energy Efficiency
and Renewable Energy (EERE) under the Hydrogen Fuel Cell Technology
Office, Award Number DE-EE0010724. The views expressed herein do not
necessarily represent the views of the U.S. Department of Energy or the
United States Government.

\section*{References}\label{references}
\addcontentsline{toc}{section}{References}

\phantomsection\label{refs}

\end{document}